\documentclass[%
	 reprint,
	preprintnumbers,
	nofootinbib,
	 amsmath,amssymb,
	 aps,
	floatfix,
	multicol,
	]{revtex4-1}
	
\usepackage[italicdiff]{physics}
\usepackage{comment}
\usepackage{graphicx,color,hyperref}
\usepackage{dcolumn}
\usepackage{bm}
\usepackage{subfigure}

\usepackage{ulem}
\usepackage{amsthm}

\def\W{{\rm W}}

\usepackage{ulem}

\usepackage{listings}
\theoremstyle{plain}

\begin{document}

\preprint{OU-HET 1154}
\preprint{KUNS-2941}

\vspace*{5mm}

\title{Deriving dilaton potential in improved holographic QCD from chiral condensate}

\author{Koji Hashimoto}%
\email{koji@scphys.kyoto-u.ac.jp}
\affiliation{%
Department of Physics, Kyoto University, Kyoto 606-8502, Japan
}
\author{Keisuke Ohashi}%
\email{keisuke084@gmail.com}
\affiliation{%
Research and Education Center for Natural Sciences, Keio University,
Kanagawa 223-8521, Japan
}
\author{Takayuki Sumimoto}%
\email{t\_sumimoto@het.phys.sci.osaka-u.ac.jp}
\affiliation{%
Department of Physics, Osaka University, Toyonaka, Osaka 560-0043, Japan
}

\begin{abstract}
We derive an explicit form of the dilaton potential in improved holographic QCD (IHQCD) 
from the QCD lattice data of the chiral condensate as a function of the quark mass.
This establishes a data-driven holographic modeling of QCD --- machine learning holographic QCD.
The modeling consists of two steps for solving inverse problems. 
The first inverse problem is to find
the emergent bulk geometry consistent with the lattice QCD simulation data at the boundary.
We solve this problem with the refinement of neural ordinary differential equation, 
a machine learning technique.
The second inverse problem is to derive a bulk gravity action with a dilaton potential 
such that its solution is the emergent bulk geometry.
We solve this problem at non-zero temperature,
and derive the explicit form of the dilaton potential.
The dilaton potential determines the bulk action, the Einstein-dilaton system, thus we derive
holographically the bulk system from the QCD chiral condensate data.
The usefulness of the model is shown in the example of the prediction of the string breaking distance, whose value is found to be consistent with another lattice QCD data.

\end{abstract}

\maketitle


\section{Introduction}
\label{sec:intro}

Machine learning is one of the major methods in theoretical 
and experimental physics.
The use is particularly effective in fields with vast amount of
data, while it is still quite limited in formal aspects
of theoretical physics.

For example in the research of AdS/CFT correspondence \cite{Maldacena:1997re,Gubser:1998bc,Witten:1998qj},
in spite of numerous pairs of the bulk and the boundary QFT
have been found, the strategy mainly uses symmetric properties
of the physical systems, and is not based on the data science.
However, let us note that if one formulates the AdS/CFT study 
as a problem of finding the bulk theory for a given boundary QFT data,
it is a data science --- the feature extraction of the
vast amount of data of the QFT to interpret it as a higher-dimensional
gravity theory. For this kind of problem, machine learning can help
to find the bulk theory.

For QCD, we already have a rich variety of lattice and experimental data,
with which we can try to solve this problem to find the holographic
bulk gravity theory dual to QCD.
The readers notice that this is an inverse problem. 
In the standard holographic modeling
of QCD, one first come up with a gravity action, and solve it to find
a metric function and a geometry. Then one put a favorite probe on it 
to compute quantities which is dual to the physical quantities in QCD,
for example, putting a fundamental string gives a Wilson loop,
or putting a probe scalar field gives a chiral condensate. 
On the other hand, what we really want in the holographic modeling of
QCD is the gravity action, for a given QCD data. Therefore, 
this is an inverse problem.

It has been proven that some of the holographic modeling has been efficiently
made by the use of machine learning \cite{Hashimoto:2018ftp,You:2017guh,Hashimoto:2018bnb,Hu:2019nea,Hashimoto:2019bih,Han:2019wue,Tan:2019czc,Hashimoto:2020jug,Yan:2020wcd,Akutagawa:2020yeo,Chen:2020dxg,Song:2020agw,Hashimoto:2021ihd,Lam:2021ugb,Katsube:2022ofz}.
In particular, the method proposed in \cite{Hashimoto:2018ftp,You:2017guh} regards the discretized bulk geometry 
as a neural network, 
and the network weights as the bulk spacetime metric.
The data of the boundary QFT is the input data of the neural network.
Thus the deep learning (DL), the machine learning using deep neural networks, 
has a similarity to the AdS/CFT correspondence,
and works as a solver of the inverse problem. 
Once the training of the neural network is made, the bulk metric is determined automatically.
This is called the AdS/DL correspondence.

Once the desired metric is obtained by the machine learning from the boundary QFT data, 
the remaining task is to find a gravity action whose solution is the metric function.
The gravity system may have functional arbitrariness. For example in the improved holographic QCD \cite{Gursoy:2007cb,Gursoy:2007er},
a dilaton-gravity system is employed as the bulk action, and the dilaton potential allows
the arbitrariness to host various QCD phenomena.
In \cite{Hashimoto:2021ihd}, we have explicitly derived a dilaton potential of a dilaton-gravity bulk system 
by requiring that its equations of motion are solved by the
metric found in \cite{Akutagawa:2020yeo}. This metric is that of the AdS/QCD model
found to be consistent with the $\rho$-meson spectral data in QCD experiments.
Therefore, the double-layer inverse problem was solved in \cite{Akutagawa:2020yeo,Hashimoto:2021ihd} 
for this particular example of the hadronic spectra at zero temperature, and the
data-driven modeling was complete for it.

In this paper, we study the non-zero temperature case to see whether this data-driven
modeling of machine-learning holographic QCD works in more generic situations. 
Our goal is to derive a dilaton potential from the data in the non-zero temperature case.
We find the dilaton potential of the bulk action, for the metric found in \cite{Hashimoto:2018bnb}.
This metric is holographically consistent with the lattice QCD data of
chiral condensate as a function of the quark mass. The solver is the neural ordinary differential 
equations (neural ODE) \cite{NODE}, a machine learning technology suitable for continuous systems
rather than discretized systems. 
The gravity equations at non-zero temperature has less symmetries: the metric has two independent
components to be determined (on the other hand, the zero-temperature metric has only one component
to be determined). Thus the determination of the bulk action by its solution is even more nontrivial.
We completely demonstrate that it is possible, and derive the dilaton potential in the bulk, 
only by using the explicit form of the metric solution.

With the obtained bulk action, we can compute the Wilson loop as a prediction of the holographic
model. The model predicts that the string breaking distance at the temperature $T=208$ [MeV] is
$d_{\rm W}^{\rm c}\sim 0.5$ [fm]. This prediction turns out to be 
consistent with various lattice QCD simulation results, which shows a nontrivial consistency 
and the predictive power of the machine-learning holographic model and the AdS/DL correspondence.

The organization of this paper is as follows. 
In Sec.~\ref{sec:NODE}, we briefly review the results obtained in \cite{Hashimoto:2018bnb} for finding
the metric from the chiral condensate data. We rerun the neural ODE with the metric ansatz
dedicated to the improved holographic QCD. In Sec.~\ref{sec:MDG} we use the metric to derive the
dilaton potential. We analyze the asymptotic behavior of the dilaton potential.
In Sec.~\ref{sec:Wilson} we calculate the holographic Wilson loop of the machine-learned model
for our prediction and find that it agrees with various other lattice QCD results.
Sec.~\ref{sec:SD} is for our conclusion and discussions.
Appendix \ref{app:detailNODE} is dedicated for the neural ODE training details, 
and Appendix \ref{sec:Z2parity}
is for the symmetric property of the fields used in this paper.


\section{New results for emergent metric in AdS/DL}
\label{sec:NODE}

The bulk geometry of the holographic QCD describing the dual of the chiral condensate
was obtained numerically in \cite{Hashimoto:2018bnb} by the use of the machine learning.
The scheme was based on the AdS/DL correspondence \cite{Hashimoto:2018ftp}.
This was refined in \cite{Hashimoto:2020jug} in which, instead of using 
deep neural networks composed of piles of layers, the neural ODE \cite{NODE} was employed
which enabled the use of continuous layers, {\it i.e.} continuous bulk geometries in
the AdS/CFT. 

In this section, we report new results which improve the bulk geometries previously
obtained in \cite{Hashimoto:2020jug}. The major improvement is about the following three respects.
 
The first improvement is about the spatial range of the bulk geometry. Originally in \cite{Hashimoto:2020jug}
the reconstructed geometry by the neural ODE was in the range $0.1\leq \eta \leq 1$ in the unit $L=1$ where 
$L$ is the AdS radius, and here we extend the range to the region  $0.01\leq \eta \leq 1$.
With this, the region near the horizon of the black hole (BH) in the bulk geometry is better approximated, as the BH horizon sits at $\eta=0$.

The second improvement is about the prior knowledge about the metric function.
The neural ODE uses a functional ansatz for the neural network weights which are bulk metric function $h(\eta)$ in our case. 
Here  $h(\eta)$ is defined  as, introducing a dilaton $\Phi$,
\begin{eqnarray}
h(\eta) \equiv  \partial_\eta \log(\sqrt{f g^3 e^{-\Phi}}),\label{eq:H}
\end{eqnarray}
in terms of a bulk metric in the  string frame 
\begin{eqnarray}
ds^2_{\rm sf}=-f(\eta)dt^2+d\eta^2+g(\eta) ds^2_{\mathbb R^3}, \label{eq:SFmetric}
\end{eqnarray}
with the 3-D flat space metric $ds^2_{\mathbb R^3}$.
{The metric \eqref{eq:SFmetric} is assumed to interpolate AdS vacuum in the limit of $\eta=\infty$ 
with the AdS radius $L$, 
\begin{align}
\lim_{\eta\to \infty }\partial_\eta \log f(\eta)=\lim_{\eta\to \infty }\partial_\eta \log g(\eta)=\frac2L  
\end{align}
and the BH horizon fixed at $\eta=0$ as 
\begin{align}
\lim_{\eta\to 0 }\eta\, \partial_\eta \log f(\eta)=2,\quad 
\lim_{\eta\to 0 }\partial_\eta \log g(\eta)=0.
\end{align}
}

{As for the machine learning applied to the emergent metric, we}
employ the following ansatz for $h(\eta)$,
\begin{align}
h(\eta) =h^{\rm nODE}(\eta)\equiv  \frac{1}{\eta} +  b_1 \eta + b_3 \eta^3 + b_5 \eta^5.
\label{hdefb}
\end{align}
The first term is the singularity due to the BH horizon, which is determined automatically for
generic horizons with non-zero temperature.
The remaining terms give our ansatz for the bulk geometry, where $b_1, b_3$ and $b_5$ are the coefficients which are trained in the neural ODE.
Note that here we have only terms with odd powers in $\eta$, while in the work \cite{Hashimoto:2020jug} all possible powers in the Taylor expansion were included. The reason of the current restriction of the powers is that in the Einstein dilaton system only the odd powers are allowed since the system 
has a certain $\mathbb Z_2$ parity.
See Appendix~\ref{sec:Z2parity} for the derivation of the restriction.

The third improvement is on the regularization. We select the form of the regularization suitable
for the Einstein-dilaton system, see Appendix \ref{app:detailNODE} for the details.

With these three improvements, we recapitulate the procedures provided in \cite{Hashimoto:2020jug}
to obtain the metric.
Table \ref{table:chiral} is the lattice QCD data which we use as the training data.
It is the data of the chiral condensate $\langle\bar{q}q\rangle[({\rm GeV})^{3}]$
as a function of the quark mass $m_{q}$[GeV] at the temperature $T=0.208$ GeV,
obtained by the RBC-Bielefeld collaboration \cite{phdthesisUnger}. 
The network architecture and the holographic model are exactly the same as that of \cite{Hashimoto:2020jug},
except for the improvement described above.
In Table \ref{table:node} we list two successfully trained results of the neural ODE.
The trained metric function is quite close to what was obtained previously in \cite{Hashimoto:2020jug,Hashimoto:2018bnb}. 
Note that 
for each of these trials in Table \ref{table:node} we slightly vary the
value of the regularization coefficients, 
therefore the difference among the trials is not the statistical error.\footnote{
For the definition of the regularization and the details of the neural network architecture, see Appendix \ref{app:detailNODE}.}

\begin{table}[t]
\centering
\begin{tabular}{c c}
$m_{q}$[GeV] & $\langle\bar{q}q\rangle[({\rm GeV})^{3}]$ \\ [0.5ex] 
\hline
\hline
0.00067 & 0.0063 \\
0.0013 & 0.012 \\
0.0027 & 0.021 \\
0.0054 & 0.038 \\
0.011 & 0.068\\
0.022 & 0.10 \\ [1ex]
\hline
\end{tabular}
\caption{Chiral condensate as a function of quark mass \cite{phdthesisUnger}, at the temperature $T=0.208$ [GeV], converted to physical units \cite{Hashimoto:2018bnb}.}
\label{table:chiral}
\end{table}

\begin{table}[t]
\centering
\begin{tabular}{c |c c |c c c}
Trial & $L$ [GeV${}^{-1}$] & $\lambda$ & $b_1$ & $b_3$ & $b_5$ \\\hline\hline
\#1 & 3.660 & 0.004868 & $-3.449$ & 6.517 & $-1.242$ \\ 
\hline
\#2 & 3.613 & 0.005587 & $-3.317$ & 6.478 & $-1.547$ 
\end{tabular}
\caption{Trained results by the neural ODE. }
\label{table:node}
\end{table}


In the following, we are going to use the trained metric \eqref{hdefb} with the coefficients given in Table \ref{table:node},
for the derivation of the dilaton potential in the dilaton-Einstein system in the holographic QCD.


\section{Emergence of Dilaton Potential}
\label{sec:MDG}

{
In this section, we determine all the components of the metric and the dilaton potential
from the metric combination $h(\eta)$ defined in \eqref{eq:H} whose explicit form was 
provided by the lattice data of the chiral condensate by the neural ODE in the previous section. 
}

{We shall determine the metric components such that these boundary conditions are satisfied
for the consistency of the AdS/CFT correspondence.}

\subsection{Einstein-Dilaton model}
In this paper, 
 we consider the five dimensional Einstein-dilaton model
\begin{equation}
S_5=\frac1{2\kappa_5}\int d^5x\sqrt{-g} \left(R-\frac43 (\partial \Phi)^2+V_D(\Phi) \right), \label{eq:EDaction}
\end{equation}
with  an unknown dilaton potential $V_D(\Phi)$\footnote{
Note that the sign of the potential is reversed from the normal one here
so that $V_D$ will be positive almost everywhere. This follows the convention of previous studies.
} which is assumed to be smooth.
The metric \eqref{eq:SFmetric} and the dilaton are supposed to be produced as a solution of this system.
{The aim of this paper is to determine the explicit dilaton potential $V_D(\Phi)$ with which the equations of
motion of \eqref{eq:EDaction} is solved with the emergent metric given in the previous section.}

{
The introduced dilaton} 
field $\Phi$ satisfies the boundary conditions
\begin{eqnarray}
\lim_{\eta \to 0} \Phi=\Phi_0,\quad \lim_{\eta \to \infty }\Phi=\Phi_\infty\equiv 0,
\end{eqnarray}
with a certain positive-definite value $\Phi_0$.
{In addition,}  
$\Phi(\eta)$ is expected to be a monotonically decreasing function,\footnote{
We guess that 
 any oscillating $\Phi(\eta)$ background implies instability  of the system against large fluctuations of $\Phi$ 
 and must be prohibited.  }
\begin{align}
\partial_\eta \Phi(\eta)<0  \quad {\rm for~} \eta>0,\label{eq:one-to-one}
\end{align}
and thus, there is a one-to-one mapping between the coordinate $\eta$ and  dilaton $\Phi$:
$\eta\in \mathbb R_{\ge 0}  \, \leftrightarrow \, \Phi=\Phi(\eta)\in (0,\Phi_0] $
as usually is in the AdS/CFT correspondence.

{Let us derive the equations of motion of the system.}
Due to the existence of dilaton,  we need to distinguish the metric in the string frame 
and that in the Einstein frame. 
{We parameterize the metric in the Einstein frame as
\begin{eqnarray}
ds^2_{\rm Ef}&=&-f_{\rm Ef} dt^2+\rho(\eta)^2 d\eta^2+g_{\rm Ef} dx_mdx^m,  \label{eq:EFmetric} \\  
f_{\rm Ef}&=&fe^{-\frac43\Phi}=\exp\left(\frac12 (\psi(\eta)+3 \chi(\eta))\right), \label{eq:deff}\\
g_{\rm Ef}&=&ge^{-\frac43 \Phi}=\exp\left(\frac12 (\psi(\eta)- \chi(\eta))\right). \label{eq:defg}
\end{eqnarray}
}
Two metrics (\ref{eq:SFmetric}), (\ref{eq:EFmetric}) are related  via a Weyl transformation,
$ds_{\rm Ef}^2=e^{-\frac43\Phi}ds^2_{\rm sf}.$
{
Then we find that the equations of motion for $\psi$, $\chi$, $\Phi$ and $\rho$ result in, respectively,
\begin{align}
0=&\,
2 \partial_\eta v+v^2+w^2+\frac43 \partial_\eta \Phi\, v 
+\frac{16}9 (\partial_\eta \Phi)^2-\frac43 V_De^{-\frac43 \Phi},\label{eq:psi}\\
0=&\, \partial_\eta w+v w+\frac23 \partial_\eta \Phi\, w,\label{eq:chi}\\
0=& \,
\partial_\eta^2 \Phi+v\,  \partial_\eta \Phi
+\frac23 (\partial_\eta \Phi)^2+\frac{3}{8}\frac{\partial V_D}{\partial \Phi}
e^{-\frac43 \Phi}\label{eq:Phi} ,\\
0=& \, 
e^{\frac{4}{3}\Phi}
\left(-\frac34 v^2+\frac34 w^2+\frac43 (\partial_\eta \Phi)^2\right)+V_D,\label{eq:rho}
\end{align}
where we 
 introduced $v\equiv \partial_\eta \psi,  w\equiv \partial_\eta \chi $, and chose a gauge
 $\rho = e^{-\frac23 \Phi}$.
 The reason for the field redefinition \eqref{eq:deff} and \eqref{eq:defg} is that the resultant equations of motion
 are simpler and diagonalized.
}

{We note here that the four equations of motion are not independent. In fact, one of them can be derived
from the other three. Below, we briefly discuss the reason for that.
Instead of introducing Eq.\eqref{eq:EFmetric}, let us write the}
metric in the Einstein frame 
as   
\begin{eqnarray}
ds^2_{\rm Ef}&=&-f_{\rm Ef} dt^2+\rho^2 d\xi^2+g_{\rm Ef} dx_mdx^m,  \label{eq:EFmetric2} 
\end{eqnarray}
where an introduced coordinate $\xi$ is related to $\eta$ with a certain one-to-one mapping,
satisfying 
\begin{equation}
d\xi=\rho^{-1} e^{-\frac23 \Phi} d\eta.
\end{equation}
By substituting metric (\ref{eq:EFmetric2}) to anction (\ref{eq:EDaction}) and omitting surface terms,
we obtain the following reduced action per unit four-dimensional volume
\begin{eqnarray}
S_{\rm rd}&=&\int d\xi \,e^{\psi} \left\{- \frac1\rho K+\rho V_D\right\}, \\
K&=&-\frac34 (\partial_\xi\psi)^2+\frac34 (\partial_\xi \chi)^2+\frac43 (\partial_\xi \Phi)^2.
\end{eqnarray}
Here $\rho$ is regarded as an einbein of this one-dimensional model and 
e.o.m of $\rho$ gives 
\begin{eqnarray}
0=\frac{\delta S_{\rm rd}}{\delta \rho}=\frac1{\rho^2} K+V_D,\label{eq:rho2}
\end{eqnarray}
which is usually regarded as a constraint. By taking this constraint into account, 
an arbitrary solution of e.o.ms of this reduced model automatically satisfies
 those for the original five-dimensional model.\footnote{
 Thanks to symmetries which the ansatz posses,  all components of the Einstein equation 
 can be written by linear combinations of those for the reduced action as
 \begin{eqnarray}
\frac{\delta S_5}{\delta g_{tt}} \propto \frac{\delta S_{\rm rd}}{\delta f_{\rm Ef}},\quad
\frac{\delta S_5}{\delta g_{\eta \eta}} \propto \frac1{2\rho}\frac{\delta S_{\rm rd}}{\delta \rho},\quad
\frac{\delta S_5}{\delta g_{mn}} \propto \frac{\delta_{mn}}{3}\frac{\delta S_{\rm rd}}{\delta g_{\rm Ef}}.
\end{eqnarray}} 
General coordinate transformation invariance  guarantees the following identity
\begin{eqnarray}
\rho \, \partial_\xi  \left(\frac{\delta S_{\rm rd}}{\delta \rho}\right)=\sum_{X=\psi,\chi,\Phi} \partial_\xi X \frac{\delta S_{\rm rd}}{\delta X}.
\end{eqnarray}
Therefore, for nontrivial configurations, 
one of three equations for $\psi,\chi, \Phi$ is automatically satisfied if the others and the constraint are solved.

{Let us obtain the consistent and independent set of equations of motion.}
By eliminating the potential $V_D$ from Eq.(\ref{eq:psi}) using Eq.(\ref{eq:rho}),
we obtain  a simpler equation
\begin{eqnarray}
0=\partial_\eta  v  + w^2 +\frac23 \partial_\eta \Phi v+\frac{16}9 (\partial_\eta \Phi)^2 \label{eq:v}.
\end{eqnarray}
We can choose  Eqs.\eqref{eq:rho}, \eqref{eq:chi} and \eqref{eq:v} as independent equations,
and  if  all of them are solved, Eq.\eqref{eq:Phi} is automatically solved 
as is mentioned above.

  {In order to solve the three equations systematically, we massage the set of equations further.}
Since  the potential $V_D$ appears only 
in Eq.\eqref{eq:rho} within those three equations, 
Eq.\eqref{eq:rho} can be regarded as an equation which  implicitly determines
the unknown $V_D$ if the one-to-one mapping \eqref{eq:one-to-one} holds. 
Since $h(\eta)$ defined in Eq.\eqref{eq:H} is calculated in terms of $v(\eta)$ as
\begin{align}
 h(\eta)=v(\eta)+\frac53 \partial_\eta \Phi(\eta), \label{eq:hv}
\end{align}
using this $h(\eta)$, $\partial_\eta \Phi(\eta)$ in Eqs.\eqref{eq:chi} and \eqref{eq:v}  can be eliminated 
and we obtain
\begin{align}
0=&\partial_\eta v+w^2+\frac{6}{25}(v-h)\left(v-\frac8{3}h\right),\label{eq:vh}\\
0=&\partial_\eta w+\frac15\left(3 v+2 h\right)w.\label{eq:wh}
\end{align}
When  a metric function $h(\eta)$ is given,   therefore, 
these two equations  are used to determine the two functions $v(\eta),w(\eta)$. 
{The field $v$ and $w$ need to} satisfy the boundary conditions
 \begin{align}
\lim_{\eta \to 0} \eta v=1,\quad \lim_{\eta \to 0} \eta w=1, \label{eq:vwbcBH} \\
\lim_{\eta \to \infty} v=\frac4L,\quad \lim_{\eta\to \infty} w=0. \label{eq:vwbcAdS}
\end{align}

{Thus, in summary, our strategy to derive the dilaton potential is as follows.
The input is the function $h(\eta)$ which the neural ODE determined explicitly. First, 
we solve Eqs.\eqref{eq:vh} and \eqref{eq:wh} under the boundary conditions \eqref{eq:vwbcBH} and \eqref{eq:vwbcAdS}, then second,
we obtain $\Phi(\eta)$ by integrating Eq.\eqref{eq:hv}, 
and finally we obtain the dilaton potential $V_D$ by using Eq.\eqref{eq:rho}.}

{As a side remark, note that} if we  set $\Phi(\eta)=\Phi_0=0$, the system reduces to one for the pure Einstein gravity and 
the exact solution for $v, w$ is given by,  with $V_D=12/L^2$,
\begin{eqnarray}
v(\eta)=v_{\rm E}(\eta)\equiv \frac{4}L \coth\frac{4\eta}L, \label{eq:vE}\\
 w(\eta)=
w_{\rm E}(\eta)\equiv \frac4L \csch\frac{4\eta}L.\label{eq:wE}
\end{eqnarray}
In this paper $\Phi(\eta)$ takes a non-trivial configuration with $\Phi_0>0$.

\subsection{Asymptotics and extrapolation of $h(\eta)$}
In Sec.\ref{sec:NODE}, $h^\text{nODE}(\eta)$ was numerically determined but it covers the limited region $0.01<\eta<1$ and it is technically hard to extend this region, whereas a solution of $v,w$ may critically depend on the boundary conditions \eqref{eq:vwbcBH} and  \eqref{eq:vwbcAdS}.
Therefore, we have to seriously consider an asymptotic behavior of $h(\eta)$ and 
need to extrapolate the trained $h(\eta)$ 
with using an appropriate extrapolation function. 
{In this subsection, we study the asymptotic regions of the spacetime: $\eta \sim 0$ and $\eta \sim \infty$.}


{
\subsubsection{Near horizon behavior at $\eta \sim 0$}
}

Focusing only on the mathematical aspects of this system of equations for $v,w$, the function $h$ can be arbitrarily imputed there, but as discussed bellow, some reasonable accompanying assumptions lead to several properties that the function must have.

At first, 
by assuming  the smoothness of
 $V_D(\Phi)$ around the point $\Phi=\Phi_0$ corresponding to the horizon $\eta=0$
we can consider  (Laurent series) expansions of $v,w$ and $h$ around $\eta=0$
and  define  $\mathbb Z_2$ parities of those functions.
All of the equations and the boundary conditions allow us to set 
 $v(\eta),w(\eta)$ and $\partial_\eta\Phi(\eta)$ to all odd functions, and in fact,
it can be shown that they must be so. See Appendix.\ref{sec:Z2parity}.
That is,   they are expanded as
\begin{align}
v(\eta)=&\frac1\eta+ \sum_{n=1}^\infty a^{(v)}_n \eta^{2n-1},\quad
w(\eta)=\frac1\eta+ \sum_{n=1}^\infty a^{(w)}_n \eta^{2n-1}, \nonumber\\
\Phi(\eta)=& \Phi_0+\sum_{n=1}^\infty a^{(\Phi)}_n \eta^{2n},
\end{align}
and especially, $h(\eta)$ must be expanded as
\begin{align}
h(\eta)=\frac1\eta+\sum_{n=1}^\infty b_{2n-1} \eta^{2n-1},
\end{align} as we have already used in Eq.\eqref{hdefb}.\footnote{
If we expect  the potential to behave 
as $V_D \sim  \Lambda e^{2Q \Phi} \Phi^P$ for large $\Phi$
with  certain constants $Q, P$ and $\Lambda$,
and the value $\Phi_0$ is also sufficiently large like ${\cal O}(10)$ even in the finite temperature,
then  $b_1$ is also expected to be
  \begin{align}
b_1 \sim  -\frac{13}{24}\left(Q-\frac{32}{39} +\frac{P}{2\Phi_0}\right) \Lambda 
e^{2( Q -\frac23) \Phi_0}\Phi_0^P.
\end{align}
According to \cite{}, they have been predicted $Q=2/3, P=1/2$ and thus,
this prediction indicates $b_1>0$, whereas its magnitude is smaller than   ${\cal O}(10)$ 
with using $\Lambda={\cal O}(10)$.   
On the other hand, if $Q\sim 2\sqrt{2}/3$, $b_1$ should take a negative value 
of much larger magnitude. 
Thus, the sign of $b_1$ has a lot to do with our expectations.
}
{Using the equations of motion, } their first terms are related with each other as
\begin{align}
 a^{(v)}_1= 2c_{\rm h}+ \tilde c_{\rm h},\quad a^{(w)}_1=-c_{\rm h} + \tilde c_{\rm h},\nonumber \\
 a^{(\Phi)}_1=-\frac94 \tilde c_{\rm h},\quad b_1=2c_{\rm h}-\frac{13}2\tilde c_{\rm h},
 \label{eq:inicoeff}
\end{align}
where $c_{\rm h}$ and $\tilde c_{\rm h}$ are defined as
\begin{align}
c_{\rm h}=&\frac29 V_D(\Phi_0)e^{-\frac43 \Phi_0},\nonumber\\
 \tilde c_{\rm h} =&\frac1{24}
\frac {\partial V_D}{\partial \Phi}(\Phi_0)e^{-\frac43 \Phi_0}.
\end{align}
{This means that}
if the functional forms of $V_D(\Phi)$ had been given, 
the coefficients would have been controlled only by the initial value $\Phi_0$
without introducing any other free parameters. 
{ In this paper, we are going backwards: we solve the equations of motion for the given $h(\eta)$ and derive $V_D$.
In the following subsection, what we will do technically is to solve the Eqs.\eqref{eq:vh} and \eqref{eq:wh} 
by a numerical shooting with varying $c_{\rm h}$. 
Another parameter $\tilde c_{\rm h}$ will be determined
automatically by the last equation in Eq.\eqref{eq:inicoeff}.
}


{
\subsubsection{Asymptotic AdS behavior at $\eta \sim \infty$}
}

Next, {let us look at the asymptotic AdS region of the spacetime.
The neural ODE provides $h(\eta)$ only in the region $\eta <L$, thus we need to find 
an appropriate extrapolation function of $h(\eta)$ for the region $\eta > L$.
For this, we study the analytic behavior of the fields in the asymptotic AdS region.}

The assumption of the AdS vacuum in the limit of $\eta=\infty$ requires that   
the potential $V_D$ behaves around the origin $\Phi=0$ as
\begin{eqnarray}
V_D(0)=\frac{12}{L^2},\quad   \frac{\partial V_D}{\partial \Phi}(0)=0.\label{eq:VDOrigin}
\end{eqnarray}
In this paper,  we just assume\footnote{Note that
the BF bound $m_\Phi^2\ge -4/L^2$ for the stability of the AdS vacuum is weaker than this condition.
Intuitively, if $0\ge m_\Phi^2\ge -4/L^2$, then long-range interaction by dilaton distorts the asymptotic AdS spacetime to the extent that it becomes an obstacle for training $h(\eta)$.
}
that the dilaton mass $m_\Phi$ around the vacuum 
satisfies
\begin{eqnarray}
m_\Phi^2\equiv -\frac{3}8  \frac{\partial^2 V_D}{\partial \Phi^2}(0) > 0.
\end{eqnarray}
Around this vacuum,  $\Phi$ and $w$  behave as
\begin{eqnarray}
\Phi\sim c_{\Phi}^+\, e^{-p_+ \eta},\quad w\sim  \frac{c_w}L\,  e^{-4 \frac\eta L},
\end{eqnarray}
and then,  $v$ 
behaves as 
\begin{align}
v  \sim & \frac 4L-\frac{8c_\Phi^+}{3L} e^{-p_+ \eta}+\frac{ c_w^2}{8L }e^{-8\frac \eta L},
\end{align}
with constants $c_\Phi^+, c_w$. 
Here $p_+$  is given by
\begin{eqnarray}
p_\pm\equiv\frac1{L} \left(2\pm \sqrt{4+m_\Phi^2L^2}\right)\in \mathbb R.
\end{eqnarray}
Therefore, $h(\eta)$ obtained in Eq.\eqref{eq:hv} must behave as
\begin{align}
h(\eta)\sim  h^{\rm ex}(\eta)\equiv  \frac4 L\left(1 - B_h e^{-p \eta} \right)\label{eq:extrah}
\end{align}
with  certain constants\footnote{{The coefficient $B_h$ controls the whole functional form of $h(\eta)$. 
In fact, when $h(\eta)$ has a valley in its profile, $B_h$ must be positive. This is
realized when }
$m_\Phi^2$ is not so large, 
\begin{align}
B_h =\frac{8+5 p_+ L}{12 } c_\Phi^+  >0,  \quad {\rm for~}
0<m_\Phi^2< \frac{32}{L^2},
\end{align}
because the assumption \eqref{eq:one-to-one}  requires $c_\Phi^+>0$.
On the other hand,  for  $m_\Phi^2> 32/L^2$, 
$B_h$ is negative as $B_h =-c_w^2/32 <0$ with $p=8/L$.}
 $B_h,p$. Under the assumption of $m_\Phi^2\ge 0$, 
$p$ is given as $p={\rm min}\{p_+, 8/L\}$ and takes a value in the narrow region\footnote{If we consider the case with $0>m_\Phi^2\ge -4/L^2$, $p$ is given by $p=p_-$ and thus satisfies
\begin{align}
0< p  \le \frac{2}L,
\end{align}
since $\Phi$ behaves as $\Phi\sim c_\Phi^- e^{-p_- \eta}$. 
In this case, there are technical difficulties to train the function $h(\eta)$ from the data in Table \ref{table:chiral}.
}
\begin{eqnarray}
\frac{4}L< p \le \frac8L. \label{eq:pineq}
\end{eqnarray}

After these analytical observations, 
we find that 
it is natural to extrapolate the trained $h(\eta)=h^{\rm nODE}(\eta)$ in Eq.\eqref{hdefb} to 
a region for large $\eta$ by using $h^{\rm ex}(\eta)$ given in Eq.\eqref{eq:extrah} as 
\begin{align}
h(\eta) =\left\{ 
\begin{array}{cc}
h^{\rm nODE}(\eta) & {\rm for~} \eta <1\\
h^{\rm ex}(\eta) & {\rm for~} \eta \ge 1  \label{eq:extrapolation}
\end{array}\right.
\end{align}
in the unit $L=1$, where two parameters $p, B_h$ are 
set by requiring the smoothness of $h(\eta)$ at $\eta=1$ as
\begin{eqnarray}
h^{\rm nODE}(1)=h^{\rm ex}(1),\quad \partial_\eta h^{\rm nODE}(1)=\partial_\eta h^{\rm ex}(1).
\end{eqnarray}

{Note that} we have to check if $p$ satisfies $4<p\le 8$.
{In fact,} many trained $h^{\rm nODE}(\eta)$ 
with poor initial conditions have violated this with deriving too large $p$ and been rejected.
{Therefore we need to include regularizations for the neural ODE training, see Appendix.\ref{app:detailNODE}.}
Finally,  
with $h^{\rm nODE}(\eta)$ given in Table \ref{table:node},  
we obtain reasonable data for $h^{\rm ex}(\eta)$.
We list it in Table \ref{table:pBh}.

\begin{table}[t]
\centering
\begin{tabular}{c |c c c c}
Trial & $p$ & $B_h$ \\\hline\hline
\#1 & 7.582 & $575.5$  \\ 
\hline
\#2 & 5.330 & $ 71.51$
\end{tabular}
\caption{{The values of $(p,B_h)$ determined to be smoothly connected to} the
neural ODE results. }
\label{table:pBh}
\end{table}

\begin{figure}[t]
\centering
\includegraphics[width=8cm]{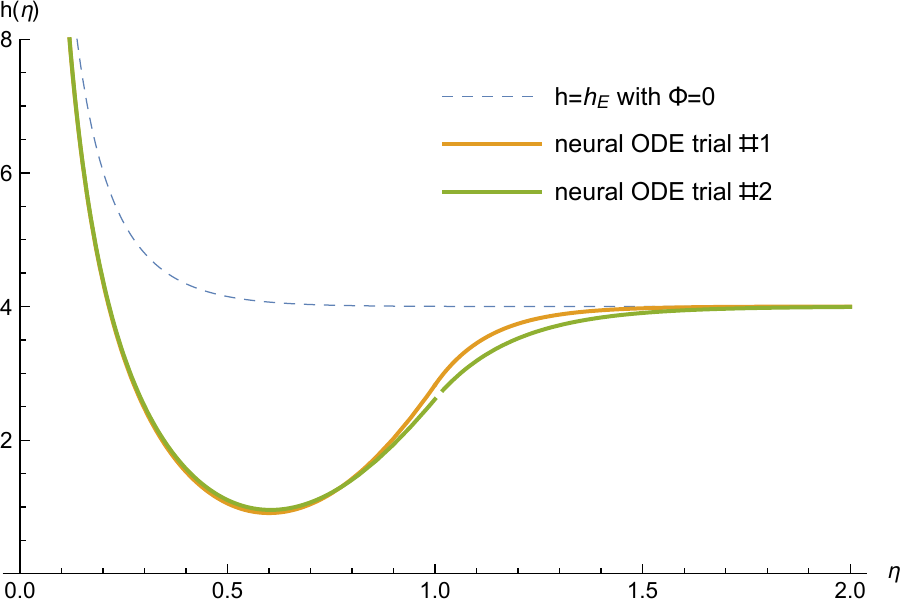}
\caption{{Fully extrapolated} profiles of $h(\eta)$. {Trial \#1 and trial \#2 are
shown,} in the unit $L=1$.
Differences between trial \#1 and trial \#2 become 
apparent in areas where the extrapolations have been used. 
{The pure Einstein gravity case $h_E(\eta)$, shown with the dashed line, has no valley.}
}\label{fig:Htrial}
\end{figure}

We show the
profiles of $h(\eta)$ for trial \#1 and trial \#2 
as thick solid lines in Fig.\ref{fig:Htrial}. 
One can see that 
there appears a deep valley in its profile.
On the other hand, in the Einstein gravity case, 
$h_{\rm E}$ is explicitly given as 
$h(\eta)=h_{\rm E}(\eta)\equiv v_{\rm E}(\eta)$ with $v_{\rm E}(\eta) $  in Eq.\eqref{eq:vE},
and is  also plotted in Fig.\ref{fig:Htrial} 
where  there is no valley in this case. 
The resulting $h(\eta)$ with the neural ODE is, therefore, 
 qualitatively different from that in the Einstein gravity case.

\subsection{Deriving  $V_D(\Phi)$ from $h(\eta)$}
\def\etair{{\eta_{\rm ir}}}
\def\etauv{{\eta_{\rm uv}}}
Here we are ready to calculate the unknown potential $V_D(\Phi)$ numerically
 from the trained $h(\eta)$ with the 
extrapolation \eqref{eq:extrapolation}. 
{This needs a numerical integration of the equations of motion by the shooting method
regarding the unknown coefficient $c_h$, as is explained below.}

\begin{figure}[t]
\centering
\includegraphics[width=8cm]{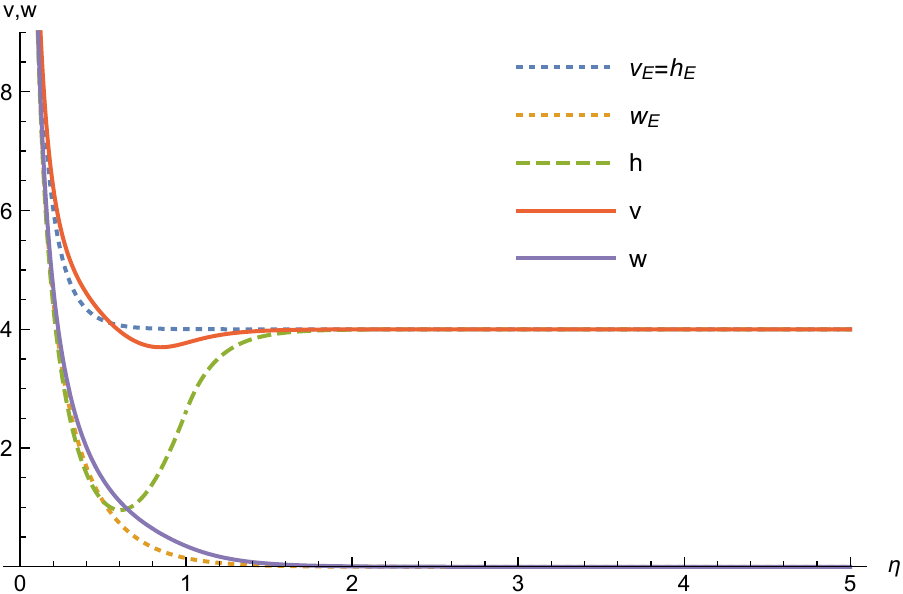}
\caption{Figures of the functions $v,w$ determined by the integration, for trial \#2.
}\label{fig:vw}
\end{figure}

Let us solve Eqs.\eqref{eq:vh} and \eqref{eq:wh} with this given $h(\eta)$ 
to obtain $v(\eta), w(\eta)$ for $\eta\in [\etair, \etauv]$ in the unit $L=1$.
To be more precise, 
we take $\etair =10^{-5}$ as {a near-horizon} 
cut-off   and 
$\etauv=5$ as {the asymptotic AdS} 
cut-off, and 
apply the shooting method  by integrating the equations from $\eta=\etair$ 
to $\eta=\etauv$.
At $\eta=\etair$, 
according to Eq.\eqref{eq:inicoeff}, 
we take  the initial values of $v, w$ as  
\begin{eqnarray}
v (\etair)=& \frac{1}\etair +\frac1{13}\left( 30 c_{\rm h} -2 b_1\right) \etair , \\
w (\etair)=& \frac{1}\etair - \frac1{13}\left(9 c_{\rm h} +2 b_1\right)  \etair,
\end{eqnarray}
where $b_1$ is a given parameter appearing in the expansion of $h(\eta)$ 
and  $c_{\rm h}$ is regarded as a free parameter since its value is unknown.
After integrating the differential equations, for large $\eta$ with $h\sim 4$, 
{the value} 
$v=h\sim 4$ turns out to be a saddle point since Eq.\eqref{eq:vh}  
reduces to be $\partial_\eta v\sim \frac85(v-4)$,
whereas $w\sim 0$ is  automatically satisfied since $w=0$ is an attractor as long as 
$3v+2 h>0$. 
Therefore we must adjust the value of $c_{\rm h}$ precisely so that $v$ can satisfy
\begin{align}
v(\etauv)=h(\etauv)\sim 4.
\end{align}
{This is the numerical shooting. We start with some ${\cal O}(1)$ value of $c_h$
and start varying the value until the equation $v(\etauv)=4$ is attained.}
The procedure determines $v(\eta), w(\eta)$ numerically as plotted in Fig.\ref{fig:vw}.
These two $v, w$ are positive definite.\footnote{
From Eqs.\eqref{eq:v} and \eqref{eq:chi} and the boundary conditions \eqref{eq:vwbcAdS}, inequalities 
can be derived as 
\begin{align}
v> \frac{4}L e^{-\frac23 \Phi},\quad w >0. 
\end{align}
}
{
The numerically searched value of $c_{\rm h}$ is
\begin{align}
c_{\rm h} = 
\left\{
\begin{array}{cc}
3.373 & \mbox{(Trial \#1)} \\
3.135 & \mbox{(Tiral \#2)}.
\end{array}
\right.
\end{align}
}

\begin{figure}[t]
\centering
\includegraphics[width=8cm]{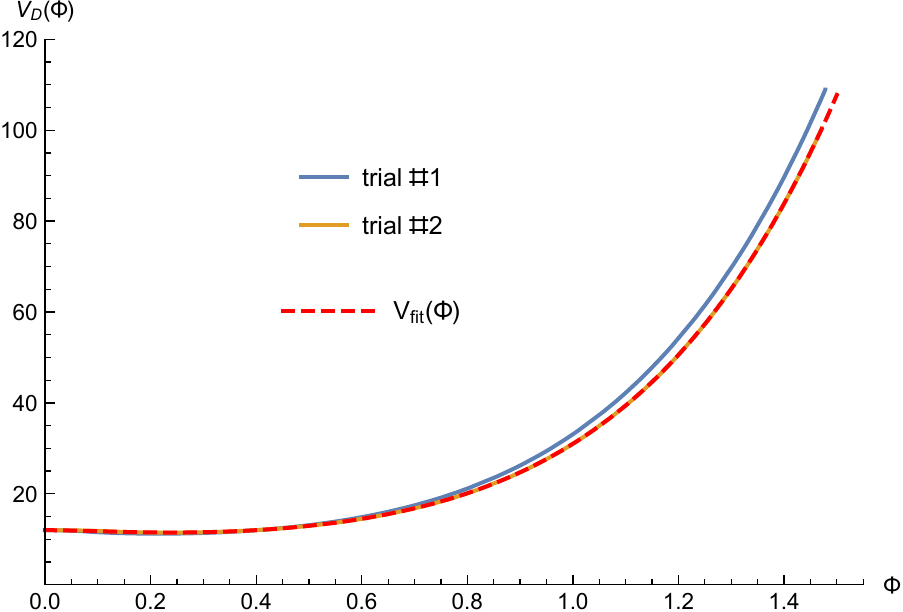}\\
\includegraphics[width=8cm]{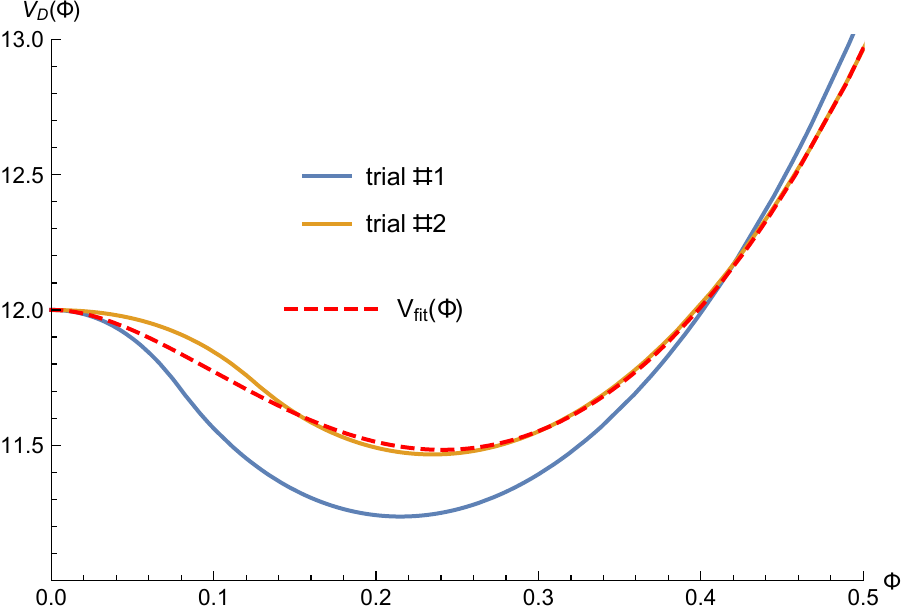}
\caption{Profiles of $V_D(\Phi)$.
The top panel shows the whole shape of the potential. 
The bottom panel is the one magnified for the vicinity of the origin.  
  }\label{fig:VDfigMS}
\end{figure}

After this procedure, thus, we can calculate the value of dilaton  by using Eq.\eqref{eq:hv} , 
\begin{eqnarray}
\Phi(\eta)&=&\frac35 \int_\eta^\etauv d\eta' (v(\eta')-h(\eta')), \label{eq:CalcPhi}
\end{eqnarray}
and  the value of the potential at $\eta$,
\begin{align}
V_D(\Phi(\eta))&=&\frac34 e^{\frac43 \Phi}\left(v^2-w^2-\frac{16}{25}(v-h)^2\right),
\end{align}
by using Eq.(\ref{eq:rho}). By combining them, finally we obtain a 
parametric representation of  the potential  $V_D(\Phi)$  
for $\Phi\in [0, \Phi_0]$ as plotted in Fig.\ref{fig:VDfigMS}.
{This concludes the complete determination of the bulk action by the QCD data of the chiral condensate, in the data-driven manner.}

{Let us study the obtained dilaton potential and discuss its physical implications.
First, observe that, }
in Fig.\ref{fig:VDfigMS}, each profile has a shallow valley.\footnote{
This is an inevitable consequence of the requirement to be $m_\Phi^2>0$.}
The resulting potential for trial \#2 is approximated by the following polynomial
except in the vicinity of  the origin of $\Phi$ corresponding to $\eta > 1$, 
\begin{align}
V_{\rm fit}(\Phi)=&12 -38.5534 \Phi^2 + 190.104 \Phi^3 \nonumber \\
&{} - 366.759 \Phi^4+ 429.313 \Phi^5 - 
    280.434 \Phi^6 \nonumber\\
    &{}+ 100.092 \Phi^7 - 14.7622 \Phi^8
\end{align}
as shown  by the dashed line in Fig.\ref{fig:VDfigMS}. 
Here, we imposed condition \eqref{eq:VDOrigin}, 
but did not force the polynomial to fit the resulting data near the origin, 
as the influence of the manual extrapolation should be dominant there.

\begin{figure}[t]
\centering
\includegraphics[width=8cm]{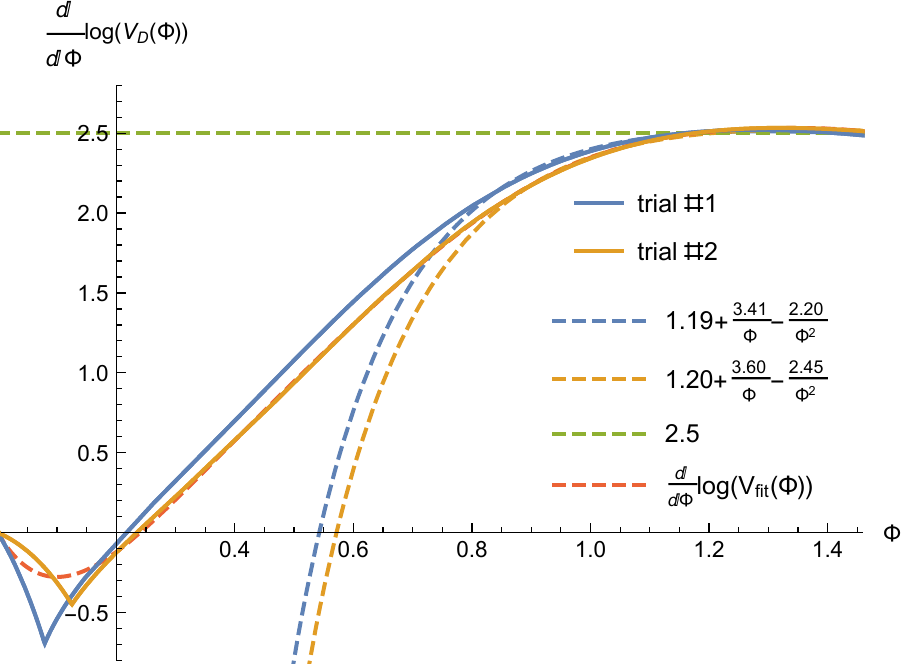}
\caption{Profiles of $\frac{d}{d\Phi}\log V_D(\Phi)$, {to extract the large $\Phi$ behavior 
of the dilaton potential $V_D$.}
  }\label{eq:VDlogdif}
\end{figure}

{We look at the large $\Phi$ behavior of the potential in more details.}
The exponent of the resulting potential for a  domain $[\Phi(0.5), \Phi(\etair)]\sim [0.85,1.5]$, that is, for relatively large $\Phi$,
can be approximated   by exponential functions as, 
\begin{align}
\frac{d}{d\Phi}\log V_D(\Phi)\sim  &
1.20+\frac{3.60}{\Phi}-\frac{2.45}{\Phi^2}.
\end{align}
as seen in Fig.\ref{eq:VDlogdif}.
{The large $\Phi$ behavior of the dilaton potential in the IHQCD is closely related to the hadron spectra.} According to \cite{Gursoy:2007cb,Gursoy:2007er},  when  $\Phi$  is sufficiently large,
the {Regge behavior of the glueball spectra leads to}
\begin{align}
\frac{d}{d\Phi}\log V_D(\Phi)\sim 2Q+\frac{P}{\Phi}
\end{align}
with  $Q=2/3$ and $P=1/2$.
We may say that our value $Q$ is close to this, and the order of our value $P$ is 
consistent with this.\footnote{
Note that
$\Phi_0=\Phi(\etair)\sim 1.5$ 
is not large, so the precise comparison needs more study.
For example,
it would be interesting to derive the glueball spectra using our dilaton potential.}

\section{Prediction: Wilson Loop}
\label{sec:Wilson}

{Since we have obtained the hull bulk action, via the AdS/CFT dictionary
we can compute various QCD quantities as a prediction of the holographic model.
Here, as one of the basic predictions, we calculate the holographic Wilson loop, 
{\it i.e.} the quark-antiquark potential,
and compare it with the lattice QCD results.}

\subsection{Temperature and metric}

The calculation of the holographic Wilson loop needs 
the metric $f,g$ in the string frame, which are given by
\begin{align}
f=e^{\frac{\psi+3\chi}2+\frac43 \Phi},\quad g=e^{\frac{\psi-\chi}2+\frac43 \Phi}.
\end{align}
{Note that in the previous section we have determined only $v = \partial_\eta  \psi$
and $w = \partial_\eta \chi$, so to get the metric components, we need the integration: }
\begin{align}
\psi(\eta)=&-\int_\eta^\infty d\eta' \left(v(\eta')-\frac4L\right)+\frac{4\eta}L+4 \delta_T,
\label{eq:vpr}\\
\chi(\eta)=&-\int_\eta^\infty d\eta' w(\eta').
\label{eq:wpr}
\end{align}
{
Here, when we make the integration, we are careful about the integration constant. The second equation \eqref{eq:wpr} does not have the integration constant since we have the boundary condition $\chi(\infty)=0$ which follows from the fact that the metric goes asymptotically to the pure AdS metric: $f/g\to 1$ as $\eta \to\infty$.
The first equation \eqref{eq:vpr} 
has a nontrivial integration constant $\delta_T$, which actually includes the temperature dependence of the metric. This constant needs to be determined such that the Hawking temperature
$T$ appearing in the standard formula
\begin{align}
f\sim 
(2\pi T \eta)^2  
\quad {\rm for ~} \quad \eta\sim 0
\label{eq:hawking} 
\end{align}
is equal to the temperature of the QCD chiral condensate data which we used for the 
machine learning, $T=208$[MeV].
Since the equation \eqref{eq:hawking} deals with the coefficient of $\eta^2$ which is difficult
to treat, let us introduce a function
\begin{align}
F(\eta)&\equiv  \frac{\psi+3\chi}4+\frac23 \Phi-\log(\tanh\frac{\eta}L)-\frac{\eta}L,
\end{align}
which behaves at the two asymptotics as non-zero constants,
\begin{align}
F(\eta) &\sim \left\{ \begin{array}{cc}
\delta_T& {\rm for}~ \eta\gg L\\
\log(2\pi T L) &{\rm for ~} \eta\sim 0.
\end{array}\right. 
\end{align}
Then using the integral expression
\begin{align}
\log(\tanh\frac{\eta}L)&=-\int_\eta^\infty d\eta'  \frac{2}L \csch\frac{2\eta'}L,
\end{align}
we find
\begin{align}
F(\eta)&=\delta_T+\frac23 \Phi
-\int^\infty_\eta \!\!\!\! d\eta' \left(\frac{v\!+\!3 w}4-\frac{2}L \csch\frac{2\eta'}L-\frac1L\right) .
\end{align}
Therefore, using the expression for $F(0)=\log(2\pi T L)$, 
we need to determine the integration constant $\delta_T$ as} 
\begin{align}
\delta_T=&\delta_{\rm ref}+\log( 2\pi TL ),\\
\delta_{\rm ref}\equiv& \int_0^\infty d\eta \left(\frac{v+3 w}4-\frac{2}L \csch\frac{2\eta}L-\frac1L\right)
-\frac23 \Phi_0.
\end{align}
{Using this formula, we can completely determine the metric components.}
Note that in reality we need to replace the integration region $(0,\infty)$
by $(\etair,\etauv)$ for our numerical purpose. 
The resulting  data are summarized in Table \ref{table:Temperature}. 

\begin{figure}[t]
\centering
\includegraphics[width=8cm]{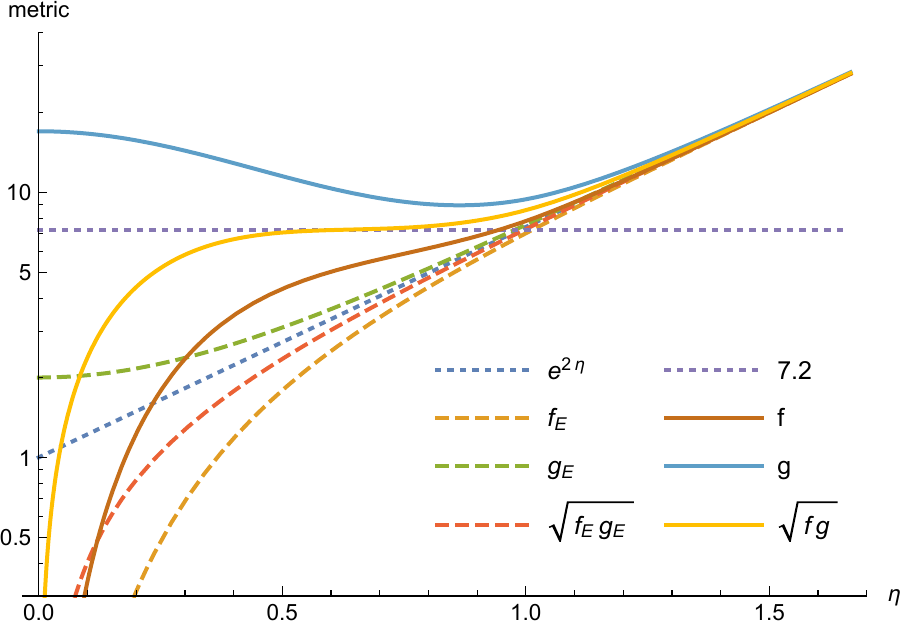}\\
\caption{
{Log plot  for the metric components $f$ and $g$  for trial \#2, and their comparison 
with the pure Einstein gravity case $f_{\rm E}$ and $g_{\rm E}$. 
We took $\delta_T=0,L=1$ for simplicity. }
}\label{fig:metrictrial}
\end{figure}

In Fig.\ref{fig:metrictrial}, we plot profiles of $f,g$ for tiral \#2  with taking $\delta_T=0$ for simplicity,
comparing them with those, $f=f_{\rm E}, g=g_{\rm E}$ calculated in the {pure}
Einstein gravity case. 
\begin{table}[t]
\centering
\begin{tabular}{c ||c | c | c |c |c }
Trial &  $TL$& $\Phi_0$ & $\delta_T$  & $d_\W^{\rm c}[{\rm fm}]$ 
& $T'_\W[{\rm GeV}^2]$\\\hline\hline
\#1 & 0.761 & 1.4782&-0.216&0.543&0.32\\
\hline
\#2 &  0.751 & 1.4718&-0.231&0.527 &0.35
\end{tabular}
\caption{Various resultant data.}
\label{table:Temperature}
\end{table}

\subsection{Holographic Wilson loop}
\begin{figure}
\centering
\includegraphics[width=8cm]{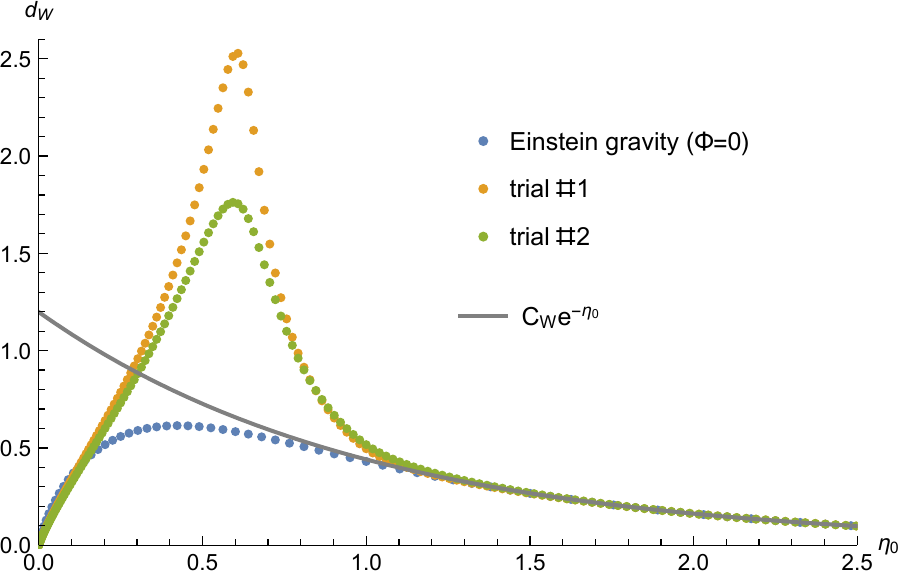}\\
\caption{Profiles of $d_\W(\eta_0)$ with 
{$L=1$. 
Here $\delta_T=0$ is formally taken for simplicity.}
}\label{fig:dtrial}
\end{figure}

Let us place a quark and an antiquark on the AdS boundary, 
keeping the {quark-antiquark separation} 
distance $d_\W$, and consider a string hanging between them,
of which the midpoint  at $\eta=\eta_0$ is  the deepest  point of the string in the radial $\eta$ direction.
 According to \cite{Maldacena:1998im,Rey:1998ik},
the distance $d_\W$ is given in terms of $\eta_0$ as
\begin{eqnarray}
d_\W=2\int_{\eta_0}^\infty \frac{d\eta}{\sqrt{g(\eta)}}\sqrt{\frac{f(\eta_0)g(\eta_0)}{f(\eta)g(\eta)-f(\eta_0)g(\eta_0)}}, 
\label{eq:dqq}
\end{eqnarray}
which is plotted in Fig.\ref{fig:dtrial}, using the training results  \#1 and  \#2.

The quark-antiquark potential $V_\W$ is calculated as 
\begin{eqnarray}
2\pi \alpha'V_\W&=&2\!\int_{\eta_0}^\infty \!\!\!\!\!
d\eta \sqrt{f(\eta)}\left(\sqrt{\frac{f(\eta)g(\eta)}{f(\eta)g(\eta)\!-\!f(\eta_0)g(\eta_0)}}-1\right)
\nonumber\\
&&\quad -2 \int_0^{\eta_0} d\eta \sqrt{f(\eta)},\label{eq:Vqq}
\end{eqnarray}
where the infinite energy of parallel two straight strings hanging from the AdS boundary  to the endpoint at the horizon has been subtracted from the potential
{to make this expression finite.}
{Since the string configuration is determined such that the free energy of the string 
is minimized, the connected string with $V_\W>0$ is not realized. If $V_\W>0$,
the configuration of the two parallel straight strings is realized. The latter means
the Debye screening of the quarks.}

\begin{figure}[t]
\centering
\includegraphics[width=8cm]{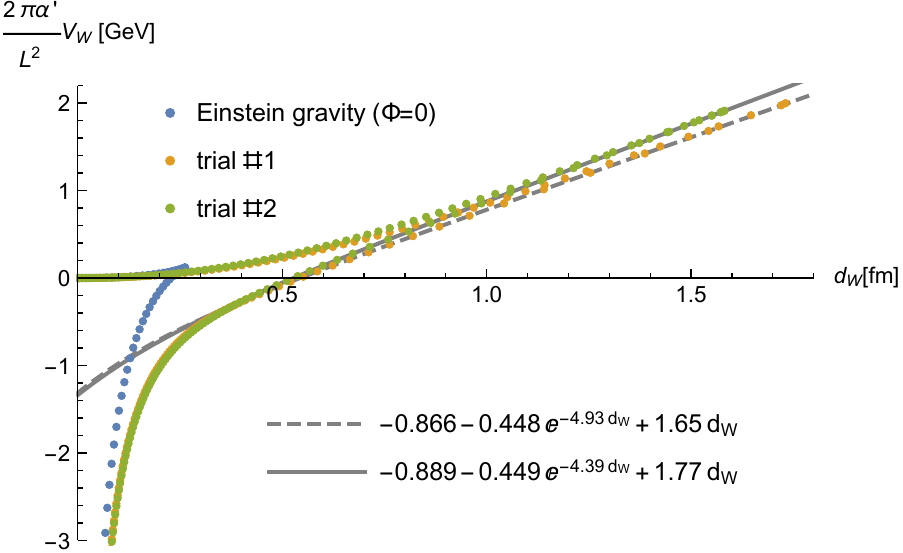}\\
\caption{Profiles of $V_\W(d_\W)$ in the Physical units. }\label{fig:Vqqtrial2}
\end{figure}

By combining the above two, we obtain a parametric representation of 
the function  $V_\W(d_\W)$ as is plotted in Fig.\ref{fig:Vqqtrial2}. 
The multivalued potential stems from the existence of two solutions 
{for $d_\W(\eta_0)$
in \eqref{eq:dqq}, see Fig.\ref{fig:dtrial}. The upper branch is never realized, so
one needs to look at only the lower branch in Fig.\ref{fig:Vqqtrial2}. 
In addition, there is another configuration which is just a set of the straight parallel strings,
giving $V_\W=0$. Thus, in Fig.\ref{fig:Vqqtrial2}, once the lower branch goes above the $d_\W$ axis, it is not realized and replaced by the line $V_\W=0$.}

These $d_\W$ and $V_\W$ 
in the physical unit {are obtained} by the following formulas
\begin{align}
d_\W=& d_\W \Big|_{L=1,\delta_T=0} \times  \frac{ e^{-\delta_{\rm ref}} }{2\pi T},\\
\frac{2\pi\alpha'}{L^2} 
V_\W=&2\pi \alpha' V_\W \Big|_{L=1,\delta_T=0} \times 2\pi T e^{\delta_{\rm ref} },
\end{align}
and  information of $e^{\delta_T}=2\pi T Le^{\delta_{\rm ref}}$ is easily restored.\footnote{
For instance, $\delta_{\rm ref}=-\log (2\sqrt{2})$ 
in the Einstein gravity case with $\Phi_0=0$.}

{Let us look at the short-distance and long-distance behavior of the obtained 
quark-antiquark potential. First, as for the short-distance behavior at small $d_\W$ 
which corresponds to $\eta_0 \gg L$, we find}
\begin{align}
\frac{2\pi \alpha'}{L^2} V_\W
= &-\frac{C_\W^2}{d_\W}+{\rm const.},
\end{align}
{ where we have used}
\begin{align}
 d_\W=&  \frac{ e^{-\delta_{\rm ref}} }{2\pi T}\times  C_W e^{-\frac{\eta_0}L},\quad C_\W\equiv\frac{(2\pi)^{\frac32}}{\Gamma(\frac14)^2}.
\end{align}
{This $1/d_\W$ behavior is typical of conformal field theories, reflecting the fact that 
the string is in the pure AdS geometry. In fact, one can check that
the quark potential is almost identical to that in the case of the pure Einstein gravity, 
see Fig.~\ref{fig:Vqqtrial2} and in particular the gray solid line in Fig.\ref{fig:dtrial}.}
{Note that the dimensionless coefficient $2\pi \alpha' L^{-2}$ is 
not the parameter of the model, so the overall normalization of the 
quark-antiquark potential cannot be predicted in this model.}

\begin{figure}[t]
\centering
\includegraphics[width=8cm]{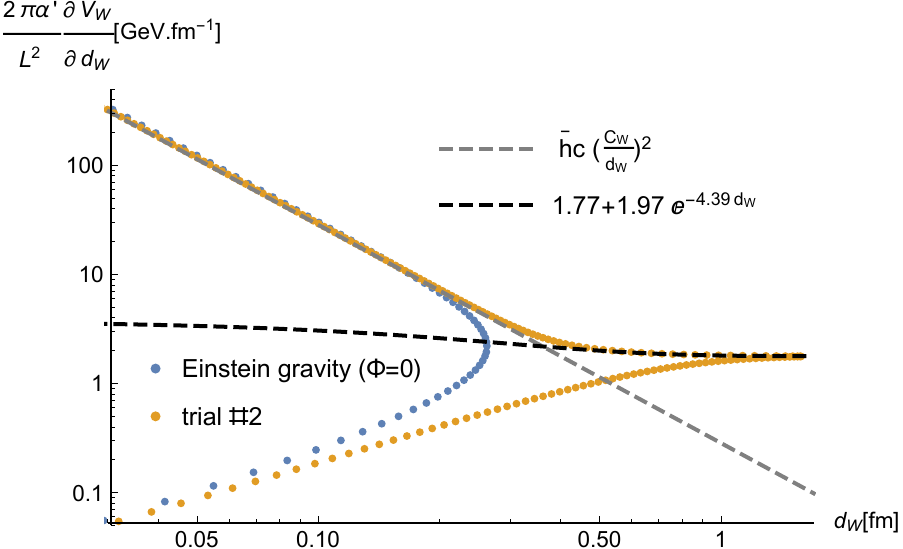}\\
\caption{Log-log plot of $\frac{\partial V_\W}{\partial d_\W}(d_\W)$  for trial \#2 in the physical units. }\label{fig:DifVqqtrial}
\end{figure}

{The long-distance behavior of the quark-antiquark potential 
differs} significantly from the case of Einstein gravity. 
{In fact, ours possesses the confining linear potential,}\footnote{
The subleading term is theoretically proven to dump exponentially.  
 See  Sec.~3.3 in \cite{Kinar:1998vq} for the proof. }
\begin{align}
V_\W\sim T_\W ( d_\W -d_\W^{\rm c})
\end{align}
with constants $T_\W$ and $d_\W^{\rm c}$, 
as seen in Fig.\ref{fig:Vqqtrial2} and Fig.\ref{fig:DifVqqtrial}.
{Therefore, the system develops the quark confinement.}
For trial\#2, the  string breaking scale $d_\W^{\rm c}$ determined\footnote{We observe that the tension $T_\W$ is 
\begin{align}
 T_\W'\equiv \frac{2\pi\alpha'}{L^2} T_\W\sim& \, 7.2 \times 
 (2\pi T e^{\delta_{\rm ref}})^2 \nonumber \\
\sim& \, 0.35 \, [{\rm GeV}^2]
=1.8 \, [{\rm GeV}\cdot {\rm fm}^{-1}]. \label{eq:TWtrial}
\end{align}
{Note that this is not the physical quantity as one needs the prefactor
$2\pi\alpha'/L^2$ to compare it with the actual string tension in QCD.}
}
as an intersection with $V_\W=0$, is
\begin{align}
d_\W^{\rm c} =&0.587\times  \frac{e^{-\delta_{\rm ref}}}{2\pi T }=2.67[{\rm GeV}^{-1}]=0.527 [{\rm fm}] .
\label{eq:pred}
\end{align}
{This breaking distance is the physical prediction of the model.}

Comparing Fig.\ref{fig:metrictrial} and Fig.\ref{fig:dtrial}, 
we find that this appearance of the linear potential is due to 
the existence of the plateau  of the metric $\sqrt{f g}$ in Fig.\ref{fig:metrictrial}.
Such a plateau does not appear in that for the Einstein gravity case.
Note that Eq.(\ref{eq:dqq}) and Eq.(\ref{eq:Vqq}) are defined for  $\eta_0$ 
so that $f(\eta)g(\eta)$ is a monotonically increasing function for all $\eta>\eta_0$
and the plateau immediately causes large value of $d_\W$ and $V_\W$. 
We can extract  this large contribution from $V_\W$ using that in $d_\W$
by rewriting Eq.\eqref{eq:Vqq} as, 
\begin{eqnarray}
2\pi\alpha'V_\W&=&
\sqrt{f(\eta_0)g(\eta_0)}\,d_\W \nonumber\\
&&{}- 2 \int_{\eta_0}^\infty d\eta \left(\sqrt{f(\eta)}-\sqrt{f(\eta)-\frac{f(\eta_0)g(\eta_0)}{g(\eta)}} \right) 
\nonumber\\
&&{}-2\int_0^{\eta_0}d\eta \sqrt{f(\eta)}.
\end{eqnarray}
In this formula, only values of $V_\W $ and $d_\W$ drastically changes in the plateau and
as a result we observe an almost linear potential there. 
We can read the tension $T_\W$ from this expression as
\begin{align}
T_\W' = \frac{2\pi\alpha'}{L^2} T_\W \sim \frac{\sqrt{f(\eta_0)g(\eta_0)}}{L^2}, \label{eq:TW}
\end{align}
which take  an almost fixed value  for $\eta_0$ in the plateau.
In fact,  we find that this formula is consistent with Eq.\eqref{eq:TWtrial}  and Fig.\ref{fig:metrictrial}.

\begin{figure}[t]
\centering
\includegraphics[width=8cm]{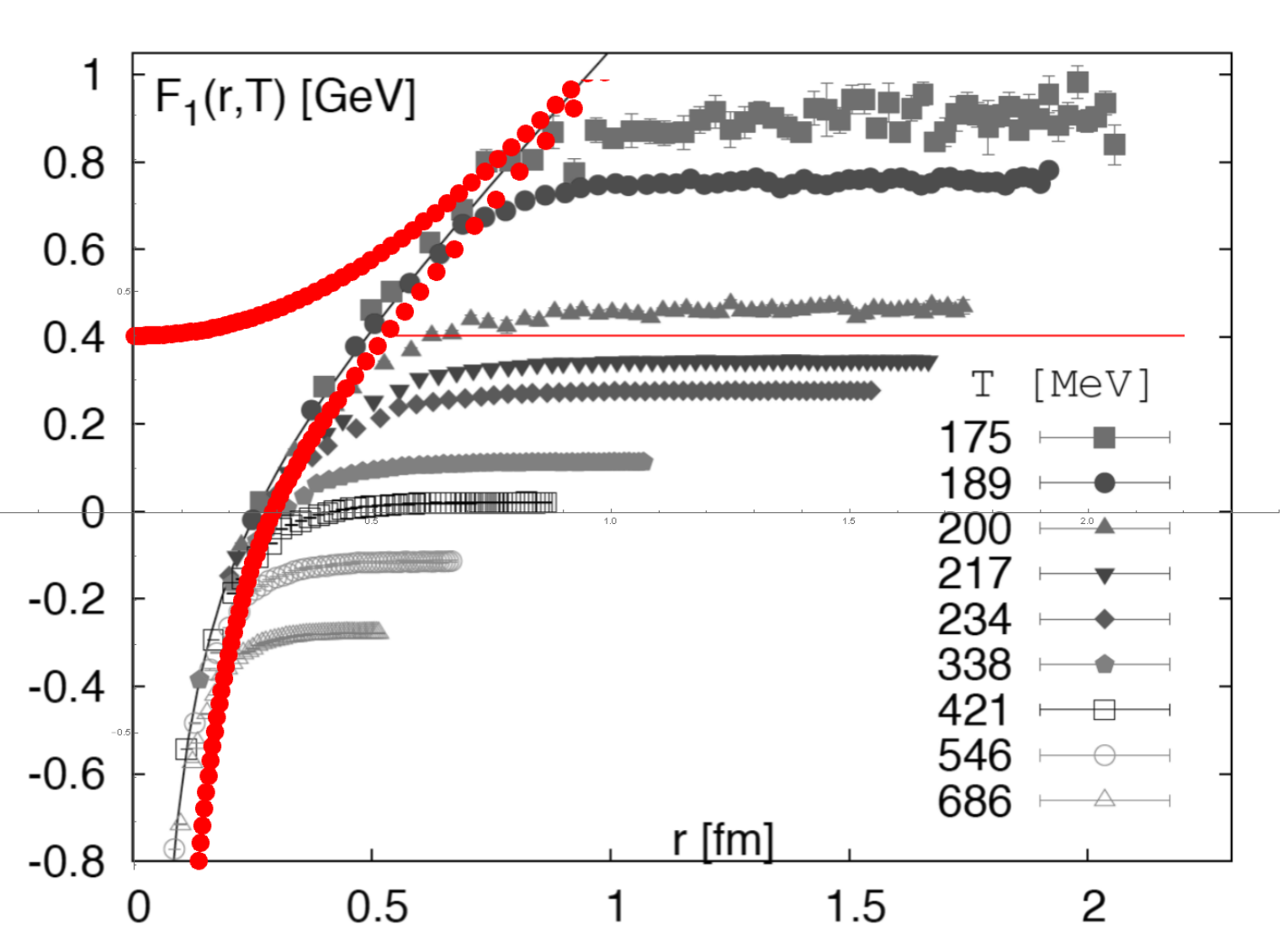}\\
\caption{
\color{black}
Comparison between the lattice data \cite{Petreczky:2010yn} (black and grey symbols) and 
our result for trial \#2 (red dots and a red straight line) for the quark antiquark potential. 
The vertical axis is the potential, and the horizontal axis is the quark separation.
The realized holographic potential is the lowest one among all the red dots and the red line
at each value of the quark distance.
To plot our result, we choose the value of 
$\alpha'$ such that the long distance behavior of our model fits that of the lattice data, to find that our fitting value is $L^2/2\pi\alpha'=0.7$.
In addition, since our model is for $T=208$ [MeV], we set the zero of our potential as $0.4$ [GeV]
in the lattice data plot, which is
the red line. }\label{fig:WilsonComparison}
\end{figure}

Finally,  in Fig.\ref{fig:WilsonComparison}
we  compare our results with those of the lattice calculation \cite{Petreczky:2010yn}, 
adjusting the origin and the scale of the vertical axis. 
{The lattice data shows that the string breaking distance at $T\sim 200$ [MeV] 
is estimated as $d \sim 0.5$ [fm], which is in good agreement with our prediction $d_\W^{\rm c}$
given in \eqref{eq:pred}.}
{This agreement ensures the validity of the present model of the emergent dilaton potential.}


\section{Summary and Discussion}
\label{sec:SD}

In this paper we have derived the dilaton potential (Fig.~\ref{fig:VDfigMS})
in improved holographic QCD, from the lattice data of the chiral condensate as
a function of the quark mass. First we have improved the neural ODE provided in \cite{Hashimoto:2020jug}
for obtaining the bulk metric from the chiral condensate data.
Then requiring that the emergent machine-learned metric is a solution of the bulk equations
of motion of an Einstein-dilaton system, we have derived the dilaton potential uniquely. 
This completes the bulk reconstruction program.

The result of the neural ODE fixes only a certain combination of the metric components.
However, requiring that they solve the equations of motion, we can constrain the
dilaton potential in the bulk action and in fact can derive it. This determines all the
metric component and the dilaton profile uniquely, at the same time.
We have used the determined metric to calculate a holographic Wilson loop, and 
have obtained a prediction of the QCD-string breaking distance $d \sim 0.5$ [fm]
at our temperature value $T=208$ [MeV] (see \eqref{eq:pred}).
This value is almost identical to what is known in lattice simulations (see Fig.~\ref{fig:WilsonComparison}).

Let us make a brief comparison between the dilaton potential obtained by the chiral condensate 
in this work and the one
obtained in \cite{Hashimoto:2021ihd} by the data of the hadron spectra. They are consistent with each other.
In fact, our dilaton potential is defined in the range of $0\leq \Phi\leq 1.5 $ while
the dilaton potential in \cite{Hashimoto:2021ihd} is in $0\leq \Phi\leq {\cal O}(20)$.
The latter is insensitive to the detailed region near the origin.
So naively we can combine these two dilaton potentials to form a consistent single dilaton potential.
This unified Einstein-dilaton system can recover both of the chiral condensate at the non-zero temperature
and the $\rho$-meson spectra at zero temperature.\footnote{
This is a qualitative statement, and still needs a detailed confirmation. For example, 
our obtained AdS radius $L\sim 3.6$ [GeV${}^{-1}$] $=0.71$ [fm] at $T=0.208$ [GeV]
is slightly different from the value $L\sim 0.51$ [fm] at $T=0$ obtained in \cite{Hashimoto:2021ihd}.
}

We should confess that it was unexpected that the string breaking distance took the right value in our model,
because the confinement and Debye screening of the quarks is not directly related to the behavior
of the chiral condensate in QCD.
In fact, in the previous estimate of the string breaking distance by the machine-learned metric
\cite{Hashimoto:2018bnb}, the estimated value is quite different from our value. There are two
reasons for that. 
First, in \cite{Hashimoto:2018bnb} a guess for a certain component of the metric was used,
while in our present work we have derived all the components of the metric by requiring
that they should be derived from a single Einstein-dilaton system. 
Another reason is that we have improved the machine learning to refine the metric.
So we conclude that the guess
of the metric component in \cite{Hashimoto:2018bnb} was too naive, and once the bulk system is specified
the prediction of the string breaking distance goes quite well.

Finally let us discuss a subtle and remaining issue of the model. In this work we considered
only the data at $T=208$ [MeV], and in the future work it is desirable to include 
all lattice data at various values of the temperature, to find a unique and consistent bulk action
from which all of the data is reproduced.
Unfortunately this appears to be difficult, as follows. 
Assuming that the dilaton potential $V_D$ does not explicitly have a temperature dependence, 
we may calculate $h(\eta)$ from the Einstein-dilaton system with the potential $V_D(\Phi)$,
and then we will find that $h(\eta)$ has no temperature dependence at all.\footnote{
Note that in this paper we have assumed $m_\Phi^2>0$ which means $\Phi=0$ is a top of a mountain 
in a sign-flipped potential. With this assumption, the dilaton profile is determined uniquely,
so there will not be any temperature dependence.
Basically this is because 
the profile of the dilaton and the initial value $\Phi_0$ are uniquely fixed
such that the radial ``motion" of the dilaton, which starts at $\Phi_0$ at ``time" $\eta = 0$, 
has to stop at the top of the sign-flipped
dilaton potential hill at `` time" $\eta = \infty$. 
Then the solutions of $v(\eta),w(\eta),\Phi(\eta)$ are uniquely determined by giving an initial value  $\Phi_0$
(see Eq.~\eqref{eq:inicoeff}), and thus the chiral condensate does not depend on
the temperature. Note that 
the explicit temperature dependence in the metric appears only as a common overall coefficient  
$e^{2\delta_T}$  of $f,g$, as
\begin{eqnarray}
(f,g)=(f,g)\Big|_{\delta_T=0}  e^{2\delta_T}  
=(f,g)\Big|_{T=T_{\rm ref}}  \times \left(\frac{T}{T_{\rm ref}}\right)^2,\qquad
\end{eqnarray}
with $T_{\rm ref}\equiv e^{-\delta_{\rm ref}}/2\pi L$. This overall factor is just the 
integration constant and does not appear in the expression of $h(\eta)$ which determines
the chiral condensate.
In this paper we have assumed $m_\Phi^2>0$,  
otherwise the strategy we have taken here does not work well technically.
On the other hand, if we have adopted 
the case $m_\Phi^2\le 0$, 
$\Phi_0$ could to be a continuous moduli of the solution and the temperature dependence can be
encoded in $\Phi_0$.
}
This contradicts the temperature dependence of the chiral condensate.
Therefore, to find the consistent model for any temperature, we may need to loosen our 
assumptions used in this paper (such as $m_\Phi^2>0$), or we may need to consider more general
bulk action with more terms. Other possibility would be to take into account the back reaction
of the probe scalar field for the chiral condensate. These kinds of the generalized models
deserve a detailed study.

Since the data-driven holographic modeling is now possible as we have demonstrated in this paper
and in \cite{Hashimoto:2021ihd}, the next task would be to find a unified holographic QCD model
which reproduces all the physical observables of QCD. For this task, we need to compare
various inversely-solved holographic models, as we have briefly tried above. Since QCD has infinite amount of
data, the unification may need more machinery of deep learning.

\begin{acknowledgments}
We would like to thank Hong-Ye Hu for discussions.
The work of K.H.~is supported in part by JSPS KAKENHI Grant Number JP22H05115, JP22H05111 and JP22H01217.
The work of T.S.~is supported in part by JSPS KAKENHI Grant No. JP20J20628.

\end{acknowledgments}

\vspace*{15mm}


\appendix

\section{Metrics trained by neural ODE}
\label{app:detailNODE}

In this appendix, we describe the regularization methods and the training methods 
of our neural ODE in more details.
The numerical data of the trained function $h$ is described in Table \ref{table:node} with the definition \eqref{hdefb}.

\begin{figure*}[tb]
\begin{center}
\subfigure[]{
\includegraphics[scale=0.65]{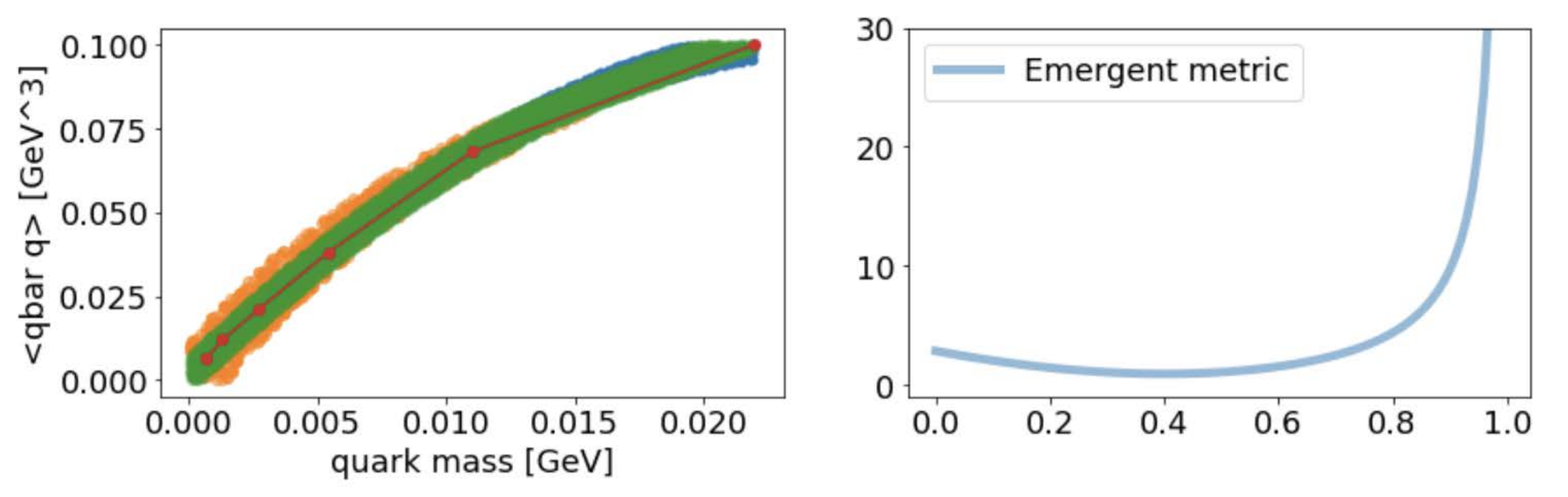}\label{Trial1}
}
\subfigure[]{
\includegraphics[scale=0.65]{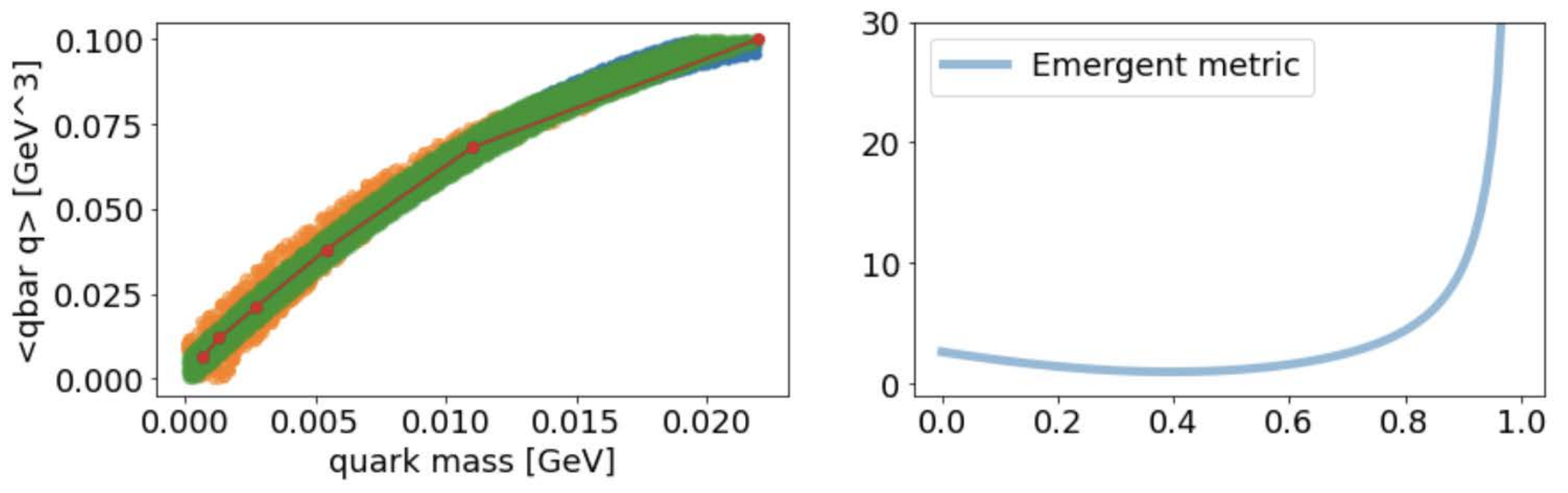}\label{Trial2}
}
    \caption{The training results of the neural ODE. (a) and (b) are for our successful trial \#1 and \#2, respectively. In each figure, the left panel shows how the QCD lattice data (shown green) is successfully reproduced by the model (orange + green). Red lines are the original central points of the lattice QCD data. The right panel in each figure shows the trained metric function $h(\tilde\eta)$ where $\tilde\eta\equiv 1-\eta$ (the region for the training is $0 \leq \eta \leq 1$ in the unit $L=1$). 
    At $\tilde\eta=1$, the BH horizon exists, while $\tilde\eta=0$ is supposed to be the AdS asymptotic region, thus approaches to $h=4$.
    }
    \label{fig:trained}
\end{center}
\end{figure*}

In Fig.~\ref{fig:trained}, we show our numerical results of the neural ODE. As the figure shows, the fit functions $h(\eta)$ reproduces the lattice QCD data well.

Let us describe our regularization method of the neural ODE. 
As mentioned in \cite{Hashimoto:2020jug}, the merit of the neural ODE is that there is no need of the smoothing 
regularization which was employed in the original AdS/DL research \cite{Hashimoto:2018ftp,Hashimoto:2018bnb} 
to regard deep neural networks to be spacetimes. 
However, still we need some regularization to make sure that
the asymptotic region of the function $h$ should approach the constant $h=4$ so that
the spacetime is asymptotically AdS${}_5$. Furthermore, since we employ Einstein-dilaton system for the holographic QCD, the emergent metric needs to approach $h=4$ in a specific manner. When the asymptotic
behavior is approximated by $h(\eta)\simeq 4-4 B_h\exp[-p\, \eta]$, the Einstein-dilaton system requires
$4<p\leq 8$ (see \eqref{eq:pineq}).

To satisfy these two kinds of the requirement, we put the following regularization terms in the loss 
function ${\cal L}$ of the neural ODE:
\begin{align}
{\cal L}_h \equiv 
c_1 (h(1)-4)^2 + c_2\left(\frac{h'(1)}{h(1)-4}+4\right)^2
\label{c1c2}
\end{align}
where $c_1$ and $c_2$ are some positive numbers called regularization coefficients 
which are hyperparameters of
the machine learning.
The first term brings the trained metric function at $\eta=1$ closer to the AdS value $h=4$, 
and the second term brings the exponent $p$ closer to the value $p=4$ which is the central value 
of the allowed region for $p$ in the Einstein-dilaton system.

During the training, we have to tune the initial conditions and 
the magnitude of the regularization for the training to be successful.
The training architecture other than the regularizations is identical to that employed in \cite{Hashimoto:2020jug}.
Our successful training is obtained by the following procedure.
First, we choose the initial conditions as follows:
\begin{align}
& L=1.0, 
\quad \lambda=0.3, \nonumber \\
&
b_1=-4.10, \quad
b_3=7.00, \quad
b_5=0.00 \, .
\end{align}
These values for $b_1$, $b_3$ and $b_5$ 
are suggested from the functional form of $h$ of the training results given in \cite{Hashimoto:2020jug,Hashimoto:2018bnb}. We choose $c_1=0.01$ and $c_2=0.0000027$ (we tune
the maximum value of $c_2$ for the training to be successful). Then after 10000 epochs of
the training, we reach
\begin{align}
& L=3.349, 
\quad \lambda=0.008769, 
\nonumber \\
&
b_1=-3.819, \quad
b_3=6.657, \quad
b_5=-0.4160 \, .
\end{align}
We use this data as another initial condition for the training with
$c_1=0.01$ and $c_2=0.001$, and after 30000 epochs, we obtain the training result \#1 given in
Table \ref{table:node}.
Finally, using the result \#1 as an initial condition, we make a training with 
$c_1=0.01$ and $c_2=0.003$ for 10000 epochs to find the training result \#2 given in Table \ref{table:node}.

In summary, the difference between the results \#1 and \#2 is the strength of the second term in the regularization \eqref{c1c2}. The training result \#2 has a value of $B_h$ closer to $p=4$, while both of \#1 
and \#2 have the values of $p$ in the consistency range $4<p\leq 8$.

{
As for the training, we also note here that 
we checked if 
the resulting  $\Phi(\eta)$ is monotonically decreasing function, 
as already declared in Eq.\eqref{eq:one-to-one}.
Or equivalently,
\begin{eqnarray}
v(\eta) > h(\eta) \quad {\rm for~}\quad \forall \eta,  \label{eq:Phi'ineq}
\end{eqnarray}
otherwise, $V_D(\Phi)$ turns out to be a multi-valued  function 
unless something miraculous happens.  
With a given trained metric  $h(\eta)$, Inequality\eqref{eq:Phi'ineq} is
 the second obstacle after Inequality\eqref{eq:pineq}  that the trained $h(\eta)$ has to  overcome.
In fact, before the modifications described in Sec.\ref{sec:NODE}, many trained data for $h(\eta)$ with  poor initial conditions,
 or a rough calculation accuracy caused 
multi-valued dilaton potential and were rejected.
}

\section{$\mathbb Z_2$ parity of functions}\label{sec:Z2parity}

Even if the dilaton potential is unknown, 
an assumption that the potential is smooth leads to a certain  $\mathbb Z_2$ parity in the solution
of $v, w$ and $\Phi$ as the followings.
Let us consider power series expansions of $v,w$ and $\Phi$ around $\eta=0$,
\begin{align}
v(\eta)=v_{\rm odd}(\eta)+ \alpha_v \eta^n+\cdots, \nonumber\\ 
w(\eta)=w_{\rm odd}(\eta)+ \alpha_w \eta^{n'}+\cdots, \nonumber\\
\Phi(\eta)=\Phi_{\rm  even}(\eta) +\alpha_\Phi \eta^{n''+1}+\cdots,
\end{align}
where $v_{\rm odd}(\eta), w_{\rm odd}(\eta)$ are odd functions with respect to $\eta$ 
and $\Phi_{\rm even}(\eta)$ is even, and the rest of terms in the r.h.s are assumed to be not so,
and  $n,n',n'' \not \in 2 \mathbb Z+1$ are assumed to give the smallest powers within them, satisfying  $n,n',n'' >-1$. 
By substituting these to the equations (\ref{eq:psi}),(\ref{eq:chi}) and (\ref{eq:Phi}) and 
taking an expansion around $\eta=0$, 
we find that contributions coming from only 
$v_{\rm odd},w_{\rm odd}$ and $\Phi_{\rm even}$ give 
expansions of even powers. And thus focusing on the lowest order terms in the equations such that their order are not even,  
those contributions must give the linearized equations
for  the terms $\alpha_v \eta^n,\alpha_w \eta^{n'}$ and $\alpha_\Phi \eta^{n''+1}$ 
that are independent of the other contributions as,
\begin{align}
0=&2(n\!+\!1) \alpha_v \eta^{n\!-\!1}+2\alpha_w \eta^{n'\!-\!1}
+\frac{4}3 (n''\!+\!1)  \alpha_\Phi  \eta^{n''\!-\!1} + \cdots\\
0=&(n'\! +\!1)\alpha_w \eta^{n' \!-\!1}+\alpha_v \eta^{n\!-\!1}
+\frac{2}3 (n''\!+\!1)  \alpha_\Phi  \eta^{n''\!-\!1} +\cdots\\
0=&(n''+1)^2\alpha_\Phi \eta^{n''-1}+2a_1^{(\Phi)}\alpha_v \eta^{n+1} +\cdots
\end{align}
with using the leading terms of $v_{\rm odd},w_{\rm odd},\Phi_{\rm even}$ as background fields,
\begin{align}
v_{\rm odd}(\eta)\sim \frac1\eta,\quad w_{\rm odd}(\eta)\sim \frac1\eta,\quad 
\Phi_{\rm even}(\eta)\sim \Phi_0+a_1^{(\Phi)}
\eta^2.
\end{align}
If $\{\alpha_v,\alpha_w,\alpha_\Phi\}$ are assumed to be non-trivial, 
the first two equations requires $n=n'=n''$ and $\alpha_v, \alpha_w \propto \alpha_\Phi$, 
whereas the last equation requires that $\alpha_\Phi $ vanishes 
and that is, $\alpha_v=\alpha_w=\alpha_\Phi=0.$
This inconsistency shows that $v(\eta), w(\eta)$ must be odd functions and $\Phi(\eta)$ are even.
That is, $f,g$    and $\Phi$  have formally the following $\mathbb Z_2$ properties,
\begin{eqnarray}
f(-\eta)=f(\eta), \quad g(-\eta)=g(\eta),\quad \Phi(-\eta)=\Phi(\eta),\quad
\end{eqnarray}
although we do not consider the inside of the BH horizon for $\eta<0$.
Especially, $h(\eta)$ must be an odd function
\begin{align}
h(-\eta)=-h(\eta).
\end{align}




\end{document}